\documentclass[aps,prd,groupedaddress,showpacs,nofootinbib]{revtex4}
\usepackage{graphicx,amssymb,amsmath}
\usepackage[english]{babel}
\usepackage{graphics}                 % Packages to allow inclusion of graphics
\usepackage{color}                    % For creating coloured text and background
\usepackage{hyperref}

\newcommand{\be}{\begin{equation}}\newcommand{\ee}{\end{equation}}
\newcommand{\bea}{\begin{eqnarray}}\newcommand{\eea}{\end{eqnarray}}
\newcommand{\beaa}{\begin{eqnarray}}\newcommand{\eeaa}{\end{eqnarray}}
\newcommand{\ba}{\begin{array}}\newcommand{\ea}{\end{array}}
\newcommand{\bit}{\begin{itemize}}\newcommand{\eit}{\end{itemize}}
\newcommand{\ben}{\begin{enumerate}}\newcommand{\een}{\end{enumerate}}

\def\lf{\left}

\def\non{\nonumber}

\def\ri{\right}

\def\1{{_{1}}}\def\2{{_{2}}}

\begin{document}

\title{Probing $CPT$ violation in meson mixing by non-cyclic phase}

\author{Antonio Capolupo }

%\vspace{2mm}

\affiliation{
Dipartimento di Matematica e Informatica,
 Universit\`a di Salerno, 84100 Salerno, Italy.}

%\maketitle

\date{\today}

\vspace{2mm}

\begin{abstract}

 The presence of non-cyclic phases is revealed in the time evolution of mixed meson systems. Such phases are related to the  parameter $z$ describing the $CPT$ violation; moreover, a non zero phase difference between particle and antiparticle arises only  in presence of  $CPT$ symmetry breaking. Thus, a completely new test for the $CPT$ invariance   can be provided by the study of such phases in mixed mesons.
 Systems which are particularly interesting for such an analysis are the $B_{s}^{0}-\bar{B}_{s}^{0}$ and the $K^{0}-\bar{K}^{0}$ ones.
 In order to introduce non-cyclic phases, some  aspects of the formalism describing the mixed neutral mesons are analyzed. Since the effective Hamiltonian  of  systems like $K^{0}-\bar{K}^{0}$, $B^{0}-\bar{B}^{0}$, $B_{s}^{0}-\bar{B}_{s}^{0}$, $D^{0}-\bar{D}^{0}$ is non-Hermitian and non-normal, it is  necessary to diagonalize it by utilizing the rules of non-Hermitian quantum mechanics.

\end{abstract}


\maketitle

\section{Introduction}

 Quantum mixing of  particles is  among the most intriguing topics in subnuclear physics.
The theoretical aspects of this  phenomenon   have been analyzed thoroughly   in the contexts of quantum mechanics (QM) \cite{Kabir}--\cite{fidecaro} and of quantum field theory (QFT) \cite{Blasone:1998hf}--\cite{Capolupo:2010ek} where modifications to the QM oscillation formulas  have been obtained. The field-theoretical corrections, due to the non–perturbative vacuum structure associated with particle mixing,  may be as large as $5 -– 20\% $  for strongly mixed systems, such as $\omega-\phi$ or for $\eta-\eta'$, \cite{CapolupoPLB2004}. On the contrary in meson systems as $K^{0}$, $B_{d}^{0}$, $B_{s}^{0}$ and $D^{0}$ and in the fermion sector these corrections are negligible.
Then, although  the QFT analysis discloses features which cannot be ignored (see for example Refs.\cite{Capolupo:2006et}--\cite{Capolupo:2010ek}), nevertheless
 a correct phenomenological description of systems as $B^{0}-\bar{B}^{0}$ can be also dealt with  in the context of QM, neglecting the non–perturbative field-theoretical effects.

The analysis of mixed meson systems  has  played  a crucial
role in the phenomenology. Indeed the  mixing  of $K^0-\bar{K}^{0}$ provided  the first
evidence of $CP$ violation in weak interactions \cite{christenson} and the $B^0-\bar{B}^{0}$
  mixing is used  to determine experimentally the precise profile of CKM
unitarity  triangle  \cite{KM}, \cite{cabibbo}. Moreover, particle mixing
offers the possibility to investigate new physics beyond the Standard Model of elementary particle physics, in particular allows to test the validity of the $CPT$ symmetry which is supposed to be an  exact symmetry.
Up to now all possible tests  are consistent with no $CPT$ violation \cite{Kostelecky:2008ts}; however, new and much more precise measurements are expected in the next generation of experiments at LHC, where $B^{0}_{s}$ and $B^{0}_{d}$ mesons will be abundantly produced  and where the very high time resolution of order of $40 fs$ of the detectors ATLAS and CMS will permit to track precisely the time evolution of the $B$ particles.

On the other hand, in recent years great attention has been devoted to the study of geometric phases \cite{Berry:1984jv}--\cite{Bruno:2011xa}  which appear in the evolution of many physical systems. Berry like phases and non-cyclic invariants  associated to neutrino oscillations (see for example \cite{Wang:2000ep}--\cite{Blasone:2009xk} and references therein) and to non-hermitian systems (see for example \cite{Mukunda}, \cite{Wang1}--\cite{Wang4} and references therein)  have been also studied extensively.

In the present paper, it is shown that these most interesting issues
are intimately bound together in such a way that
 the non-cyclic phases for mixed meson systems appear to provide a new instrument to test the $CPT$ symmetry.
It is shown that
 phases such as the Mukunda--Simon ones \cite{Mukunda}, appearing as observable characterization of the  mixed mesons evolution, are related to the parameter  denoting the $CPT$ violation. In particular, the presence of non-trivial Mukunda--Simon phases and of a phase difference between particle and antiparticle indicates unequivocally the $CPT$ symmetry breaking in mixed boson systems. Furthermore, it is shown that the non-cyclic phases can be useful also to analyze the $CP$ violation.
An  especially interesting system for studying the geometric phases is the $B_{s}^{0}-\bar{B}_{s}^{0}$ one because a lot of particle-antiparticle oscillations occur within its lifetime.
 Thus, the next experiments on the  $B_{s}^{0}$ mesons might open new horizons to be explored in future research.

In Appendix C, the Aharonov--Anandan invariants \cite{Anandan:1990fq} for mixed mesons are presented   and  their relation with the parameter describing $CP$ violation is shown.

In order to study the non-cyclic phases, some of the features of the formalism depicting the evolution of mixed neutral mesons in QM have been analyzed. Since   in the Wigner-Weisskopf approximation \cite{Wigner} the effective Hamiltonian describing such  systems is non-Hermitian and non-normal, to diagonalize it the rules of non-Hermitian quantum mechanics have to be used. In this work, the biorthonormal basis formalism \cite{Bender98}--\cite{Dattoli} will be used.

The structure of the paper is the following: in Section II the effective Hamiltonian $\mathcal{H}$ of mixed meson systems is diagonalized by a complete biorthonormal set of  states. The Mukunda--Simon phases for mixed mesons, their connections with   $CPT$ and $CP $ violations and the analysis of such phases for $B_{s}$  mesons  are presented in Section III. Section IV is devoted to the conclusions.

Useful expressions of  the states $|M^{0}(t)\rangle$ and $|\bar{M}^{0}(t)\rangle$ are reported in Appendix A. In Appendix B,  the asymmetries describing the $T$ and $CPT$ violations are computed using the biorthonormal basis formalism. They coincide with the corresponding ones obtained by employing the states usually adopted in the literature. The Aharonov--Anandan invariants are studied in Appendix C.

\section{Meson mixing and biorthonormal basis}

The time evolution of a beam of neutral boson system and of its decay products can be described as
$
|\psi (t)\rangle \,=\, \psi_{M^{0}} (t) |M^{0} \rangle \,+\,\psi_{\bar{M}^{0}} (t) |\bar{M}^{0} \rangle \,
+\,\sum_{n} d_{n}(t)|n\rangle\,,
$
where $M^{0}$ denotes $K^{0}$, $B_{d}^{0}$, $B_{s}^{0}$ or $D^{0}$; $\bar{M}^{0}$ the corresponding antiparticles,   $|n\rangle$ are
the decay products, $t $ is the proper time, $\psi_{M^{0}} (t)$, $\psi_{\bar{M}^{0}} (t)$ and $d_{n}(t)$
are time dependent functions. Since the decay products are absent at the instant of the $M^{0}$ and $\bar{M}^{0}$ production,
the state vector at initial time $t=0$ is given by
$
|\psi (0)\rangle \,=\, \psi_{M^{0}} (0) |M^{0} \rangle \,+\,\psi_{\bar{M}^{0}} (0) |\bar{M}^{0} \rangle \,.
$

If one is interested in evaluating only the wave functions $\psi_{M^{0}} (t)$ and $\psi_{\bar{M}^{0}}(t)$ and the times considered are much larger than the typical time scale of the strong interaction, then the time evolution of $|\psi (t)\rangle $ can be well described, in the space formed by $|M^{0} \rangle$ and $ |\bar{M}^{0} \rangle $, by the Wigner-Weisskopf approximation \cite{Wigner}. The time evolution is thus determined by  the Schrodinger equation
$i \frac{d }{dt} \Psi   \,=\,\mathcal{H}\, \Psi  \,,$
where $\Psi = \lf(\psi_{M^{0}}(t)\,, \psi_{\bar{M}^{0}} (t) \ri)^{T}$ and  the effective Hamiltonian $ \mathcal{H} = \left(\begin{array}{cc}
                  \mathcal{H}_{11} & \mathcal{H}_{12} \\
                  \mathcal{H}_{21} & \mathcal{H}_{22} \\
                \end{array}\right)$
of the system is non-Hermitian. It can be written as $\mathcal{H} = M - i \frac{\Gamma}{2}$ with $M$ and $\Gamma$ Hermitian matrices.
The matrix elements of $\mathcal{H}$ are constrained by the conservation of discrete symmetries \cite{fidecaro}:
$CPT$ conservation implies $\mathcal{H}_{11} = \mathcal{H}_{22}$, $T$ conservation entails $|\mathcal{H}_{12}| = |\mathcal{H}_{21}|$ and $CP$ conservation  requires
$\mathcal{H}_{11} = \mathcal{H}_{22}$ and   $|\mathcal{H}_{12}| = |\mathcal{H}_{21}|$.

Notice that  in the presence of $CP$ violation, i.e. for $|\mathcal{H}_{12}| \neq |\mathcal{H}_{21}|$, the mass matrix $M$ and the decay matrix $\Gamma$ do not commute,  $[M , \Gamma] \neq 0$, then the Hamiltonian  $\mathcal{H}$ is non-Hermitian, $\mathcal{H} \neq \mathcal{H}^{\dag}$ and non-normal,
$[\mathcal{H} , \mathcal{H}^{\dag}] \neq 0$. In this case, the left and right eigenstates of  $\mathcal{H}$ are independent sets of vectors
that are not connected by complex conjugation. This implies that $\mathcal{H}$ cannot be diagonalized by a single unitary transformation but one has to use the rules of non-Hermitian quantum mechanics.
In the following, the biorthonormal basis formalism \cite{Bender98} - \cite{Dattoli}  will be used
 and the  discussion presented in Ref.\cite{Dattoli} will be applied to describe the time evolution of mixed mesons in the presence of $CP$ violation (see also Ref.\cite{AlvarezGaume:1998yr}).

Let   $\lambda_{j}  = m_{j} - i \Gamma_{j}/2$, with $j=L,H$ ($L$ denotes the light mass state and $H$ the heavy mass state\footnote{For $K$ mesons, usually the mass eigenstates are defined according to their lifetimes: $K_S$ is the short lived and $K_L$ is the long lived; in this system $K_L$ is the heavier state.}), be the eigenvalues of the Hamiltonian $\mathcal{H}$,
with $|M_{j}\rangle $  the corresponding eigenvectors:
\bea\label{autovaH}
\mathcal{H} |M_{j}\rangle\, =\,  \lambda_{j}  |M_{j}\rangle \,.
\eea
Denoting with $\varepsilon_{j}$ and $|\widetilde{M}_{j}\rangle $, $(j=L,H)$ the eigenvalues and the eigenvectors of $\mathcal{H}^{\dag}$:
$\label{autovaHdag}
\mathcal{H}^{\dag}|\widetilde{M}_{j} \rangle  \, =\,  \varepsilon_{j} |\widetilde{M}_{j}\rangle  \,;
$
this equation can be recast in the form
\bea\label{mtilde}
\langle\widetilde{M}_{j} | \mathcal{H}  \, =\,   \langle\widetilde{M}_{j} |\varepsilon^{*}_{j}\,.
\eea
By projecting Eq.(\ref{mtilde}) on the state $|M_{j}\rangle$  one has
$\label{aspec}
\langle\widetilde{M}_{j} | \mathcal{H} |M_{j}\rangle \, =\, \langle\widetilde{M}_{j} |\varepsilon^{*}_{j}|M_{j}\rangle\,=\,
\langle\widetilde{M}_{j} |\lambda_{j}|M_{j}\rangle\,,
$
then $\varepsilon^{*}_{j}= \lambda_{j}$, i.e. the eigenvalues of $\mathcal{H}$ are the complex conjugates of those of $\mathcal{H}^{\dag}$. Moreover one has
$
\langle\widetilde{M}_{i} | \mathcal{H} |M_{j}\rangle \, =\, \langle\widetilde{M}_{i} |\varepsilon^{*}_{i}|M_{j}\rangle\,=\,
\langle\widetilde{M}_{i} |\lambda_{j}|M_{j}\rangle\,,
$
hence:
$
(\lambda_{j}-\varepsilon^{*}_{i})\langle\widetilde{M}_{i} |M_{j}\rangle\,=0\,.$
This last relation together with   $\lambda_{j} \neq \varepsilon^{*}_{i}$ for $i \neq j$  implies the biorthogonality relation
\bea\label{biorthog}
\langle\widetilde{M}_{i} |M_{j}\rangle\,= \langle M_{j}|\widetilde{M}_{i}\rangle\,=\,\delta_{ij} \,.
\eea
Let us now derive the completeness relation. The state vector $|\psi(t)\rangle$ of the neutral boson system (without its decay products)  can be expressed as
$
|\psi(t)\rangle \,=\,\sum_{j=1,2} a_{j}(t) |M_{j}\rangle \, =\,\sum_{j=1,2} \widetilde{a}_{j}(t) |\widetilde{M}_{j}\rangle\,,
$
with
$
a_{j}(t) = \langle\widetilde{M}_{j} |\psi(t)\rangle $ and $ \widetilde{a}_{j}(t) = \langle M_{j}|\psi(t)\rangle \,,
$
i.e.
$
|\psi(t)\rangle \,=\,\sum_{j=1,2}  |M_{j}\rangle \langle\widetilde{M}_{j} |\psi(t)\rangle \, =\,\sum_{j=1,2}   |\widetilde{M}_{j}\rangle
\langle M_{j}|\psi(t)\rangle\,.
$
This last equation implies the completeness relations
\bea\label{complet}
\sum_{j } |M_{j}\rangle \langle\widetilde{M}_{j}| =\sum_{j } |\widetilde{M}_{j}\rangle \langle M_{j}| =1\,.
\eea

 Summarizing, since in the presence of $CP$ violation the effective Hamiltonian $\mathcal{H}$ of mixed meson systems is non-Hermitian and non-normal, then the conjugate states $\langle\widetilde{M}_{j}|^{\dag} \equiv |\widetilde{M}_{j}\rangle$ and $| {M}_{j}\rangle^{\dag} \equiv \langle {M}_{j}|$  are not isomorphic to their duals:   $|\widetilde{M}_{j}\rangle \neq |{M}_{j}\rangle$
   and
 $\langle {M}_{j}| \neq \langle\widetilde{M}_{j} |$. In this case, as a consequence of Eqs.(\ref{biorthog}) and (\ref{complet}), the set of states $ \{|M_{j}\rangle ,  \langle\widetilde{M}_{j}| \}$, with $j=L,H$,  is a complete biorthonormal system for $\mathcal{H}$.\footnote{Note that Eqs.(\ref{autovaH}), (\ref{mtilde}),  (\ref{biorthog}) and (\ref{complet}) do not determine the biorthonormal system
$\{|M_{j}\rangle\,\,,|\widetilde{M}_{j}\rangle  \}$ uniquely. Any other biorthonormal system,
$\{|M^{\prime}_{j}\rangle\,\,,|\widetilde{M^{\prime}}_{j}\rangle  \}$, satisfying these conditions has the form
$
|M^{\prime}_{j}\rangle\,=\,\alpha_{j}|M_{j}\rangle\,$ and
$ |\widetilde{M^{\prime}}_{j}\rangle\,=\,\frac{1}{\alpha^{*}_{j}}|\widetilde{M}_{j}\rangle\,,
$
with $\alpha_{j}$ complex. This fact however does not affect any measurable quantity \cite{AlvarezGaume:1998yr}.}

Furthermore, since the time evolution operator associated with $\mathcal{H}$, $U(t)= e^{-i \mathcal{H} t}$  is not unitary, one also introduces the time evolution operator of $\mathcal{H}^{\dag}$, $\widetilde{U}(t)= e^{-i \mathcal{H}^{\dag} t}$, which satisfies $U \widetilde{U}^{\dag} =
\widetilde{U}^{\dag} U = 1$. The spectral form of the Hamiltonian and of the operators $U(t)$ and $\widetilde{U}(t)$ are then given by
\bea\label{evolution}
\mathcal{H} \,= \, \sum_{j=L,H} \lambda_{j} |M_{j}\rangle \langle\widetilde{M}_{j}|\,,
\qquad
U(t)  \,= \, \sum_{j=L,H} e^{- i \lambda_{j}t} |M_{j}\rangle \langle\widetilde{M}_{j}|\,,
\qquad
\widetilde{U}(t)  \,= \, \sum_{j=L,H} e^{- i \lambda^{*}_{j}t} |\widetilde{M}_{j}\rangle \langle M_{j}|\,,
\eea
respectively. Thus the time evolved of the states $|M_{k} \rangle$ and $|\widetilde{M}_{k} \rangle$, $(k=L,H)$ at time $t$ are
$|M_{k}(t)\rangle = U(t) |M_{k}\rangle = e^{- i \lambda_{k}t}|M_{k} \rangle $ and $|\widetilde{M}_{k}(t)\rangle = \widetilde{U}(t) |\widetilde{M}_{k} \rangle = e^{- i \lambda^{*}_{k}t} |\widetilde{M}_{k} \rangle$ and  the corresponding conjugate states are
$\langle M_{k}(t)| = \langle M_{k}| U^{\dag}(t) = \langle M_{k}|  e^{  i \lambda_{k}^{*} t} $ and $\langle \widetilde{M}_{k}(t)| = \langle\widetilde{M}_{k}|  \widetilde{U}^{\dag}(t)  = \langle\widetilde{M}_{k}|  e^{  i \lambda_{k}t}  $.

Note that the existence of a complete biorthonormal set of eigenvector of  $\mathcal{H}$  implies that $\mathcal{H}$ is diagonalizable. Thus, a matrix $V$  exists such that  $V^{-1} H V = diag (\lambda_{L},\lambda_{H})$, with
\bea
V\,=\,\left(
        \begin{array}{cc}
          p_{L} & p_{H} \\
          q_{L} & -q_{H} \\
        \end{array}
      \right)\,,
      \qquad \qquad
 V^{-1}\,=\,     \frac{1}{q_{L} p_{H}+q_{H} p_{L}} \left(
        \begin{array}{cc}
         q_{H} & p_{H} \\
          q_{L} & -p_{L} \\
        \end{array}
      \right)\,,
\eea
where $q_{L}=c_{L} $, $p_{L} = c_{L} \lf( \frac{\lambda_{L}-\mathcal{H}_{22}}{\mathcal{H}_{21}}\ri)$, $q_{H}= -c_{H} $,
$p_{H}=c_{H} \lf( \frac{\lambda_{H}-\mathcal{H}_{22}}{\mathcal{H}_{21}}\ri)$ and $c_{L} $, $c_{H} $ are complex constants: $c_{L} ,  c_{H} \in \mathbb{C}-\{0\}$.
The right and left eigenvectors of the Hamiltonian $\mathcal{H}$  are $(p_{L},q_{L})^{T}$, $(p_{H},-q_{H})^{T}$
and $ \frac{1}{q_{L} p_{H} + q_{H} p_{L}}(q_{H},p_{H})$, $ \frac{1}{q_{L} p_{H} + q_{H} p_{L}}(q_{L},-p_{L})$
respectively.\footnote{If we impose the normalization conditions of $|M_{j}\rangle$, $j =1,2,$ we have $|c_{L}| =
\frac{|\mathcal{H}_{21}|}{\sqrt{|\lambda_{L}-\mathcal{H}_{22}|^{2}+|\mathcal{H}_{21}|^{2}}}$, $|c_{H}|=\frac{|\mathcal{H}_{21}|}{\sqrt{|\mathcal{H}_{22}-\lambda_{H}|^{2}+|\mathcal{H}_{21}|^{2}}}$ and
$|p_{L}|^{2}+|q_{L}|^{2}=|p_{H}|^{2}+|q_{H}|^{2}=1$.}
 The explicit expressions of the mass eigenstates $|M_{L}\rangle$, $| M_{H}\rangle$, $\langle \widetilde{M}_{L}|$  and $\langle \widetilde{M}_{H}|$ in terms of the parameters $q_{j}$ and $p_{j}$, $j=L,H$ are given in Appendix A.
It is now convenient to introduce the $CP$ and $CPT$ parameters and to express the meson states in terms of these parameters. The constraints on $\mathcal{H}$ imposed by $CP$ and $T$ invariance suggest to adopt the following $CP$ and $T$ violation parameter:
\bea\label{CPpar}
\varepsilon =\frac{|p_{H}/q_{H}|-|q_{L}/p_{L}|}{|p_{H}/q_{H}|+|q_{L}/p_{L}|}=\frac{|p/q|-|q/p|}{|p/q|+|q/p|}= \frac{|\mathcal{H}_{12}|-|\mathcal{H}_{21}|}{|\mathcal{H}_{12}|+|\mathcal{H}_{21}|}\,,
\eea
where
\bea\label{q/p}
\frac{q}{p}\,=\,\sqrt{\frac{q_{L} q_{H}}{p_{L} p_{H}}} \,=\,\sqrt{\frac{\mathcal{H}_{21}}{\mathcal{H}_{12}}}\,.
\eea
Moreover, $CPT$ invariance imposes  the equality of the diagonal elements of the Hamiltonian $\mathcal{H}$, $\mathcal{H}_{11} = \mathcal{H}_{22} $. Thus such an invariance can be tested by checking that the difference $\mathcal{H}_{22}-\mathcal{H}_{11}$ is equal to zero. $CPT$ violation can be described conveniently by the quantity $z$ which is independent of phase convention
\bea\label{zeta}
 z\,=\, \frac{\frac{q_{L}}{p_{L}}-\frac{q_{H}}{p_{H}}}{\frac{q_{L}}{p_{L}}+\frac{q_{H}}{p_{H}}}\,
 =\,\frac{(\mathcal{H}_{22}-\mathcal{H}_{11})}{\lambda_{L} - \lambda_{H}}\,.
 \eea
Notice that in the standard model extension  (SME) the parameter $z$ depends on the four-momentum of the meson \cite{Kost}, moreover, in the case of $CPT$ invariance: $p/q = p_{L}/q_{L}=p_{H}/q_{H}$ and $z =0$.

By using Eqs.(\ref{q/p}) and (\ref{zeta}),
the mass eigenstates $|M_{L}\rangle$  and $| M_{H}\rangle$  can be written in terms of $|M^{0}\rangle$ and
$|\bar{M}^{0}\rangle$ as
\bea\label{Pa1}
|M_{L}\rangle &=& p\,\sqrt{1-z}\,|M^{0} \rangle\,+\, q\,\sqrt{1+z}\,  |\bar{M}^{0} \rangle \,,
\\\label{Pa2}
| M_{H}\rangle &=& p\,\sqrt{1+z}\,|M^{0} \rangle\,-\, q\,\sqrt{1-z}\, |\bar{M}^{0} \rangle \,,
\eea
and, in a similar way, $\langle \widetilde{M}_{L}|$  and $\langle \widetilde{M}_{H}|$  are expressed as
\bea
\label{Pa3}
\langle\widetilde{M}_{L} |  &=& \frac{1}{2 p q} \lf[q\,\sqrt{1-z}\,\langle \widetilde{M^{0}}|  \,+\,
p\,\sqrt{1+z}\,\langle \widetilde{ \bar{M}^{0}}| \ri]\,,
\\\label{Pa4}
\langle\widetilde{M}_{H}| &=& \frac{1}{2 p q } \lf[q\,\sqrt{1+z}\,\langle \widetilde{M^{0}}|  \,-\,
p\,\sqrt{1-z}\,\langle \widetilde{ \bar{M}^{0}}| \ri]\,.
\eea
Thus, at time $t$, the  states $|M^{0}(t)\rangle$ and $|\bar{M}^{0}(t)\rangle$ in terms of $|M_{L}\rangle$ and
$|M_{H}\rangle$ are
\bea\label{B0states1}
|M^{0}(t)\rangle &=& \frac{1}{2 p}\lf[\sqrt{1-z}\, |M_{L}\rangle\, e^{-i \lambda_{L}t}\,+\,
\sqrt{1+z}\, |M_{H}\rangle\, e^{-i \lambda_{H}t}\ri]\,,
\\\label{B0states2}
|\bar{M}^{0}(t)\rangle &=& \frac{1}{2 q}\lf[\sqrt{1+z}\, |M_{L}\rangle\, e^{-i \lambda_{L}t}\,-\,
\sqrt{1-z}\, |M_{H}\rangle\, e^{-i \lambda_{H}t}\ri]\,,
\\\label{B0states3}
\langle \widetilde{M^{0}}(t)| &=& p\,\lf[ \sqrt{1-z}\, \langle \widetilde{M_{L} }| \, e^{ i \lambda_{L}t}\,+\,
\sqrt{1+z}\, \langle \widetilde{M_{H} }|\, e^{i \lambda_{H}t}\ri]\,,
\\\label{B0states4}
\langle \widetilde{\bar{M}^{0}}(t)| &=& q\,\lf[ \sqrt{1+z}\, \langle \widetilde{M_{L} }| \, e^{ i \lambda_{L}t}\,-\,
\sqrt{1-z}\, \langle \widetilde{M_{H} }|\, e^{i \lambda_{H}t}\ri]\,.
\eea
The states in Eqs.(\ref{B0states1})-(\ref{B0states4}) are the correct ones to be used in computations.

\section{Meson mixing and Mukunda--Simon phase}

The main result of the paper is presented in this Section;
the Mukunda--Simon phases appearing in the time evolution of  mixed meson  $M^{0}-\bar{M}^{0}$ systems  are related to  the parameter describing the $CPT$ violation and a difference between  the non-cyclic phases of particles and of antiparticles signals $CPT$ symmetry breaking.  Moreover, the non-cyclic phases due to the particle-antiparticle oscillations are related to the parameters denoting the $CP$ violation.
A system particularly appropriate to study such phases is the $B^{0}_{s}-\bar{B}^{0}_{s}$ one.

Let us start by introducing the Mukunda--Simon phases for Hermitian systems. Subsequently, we consider such phases in the case of non-Hermitian systems and of mixed mesons.

 Consider a quantum system whose state vector $|\psi(t)\rangle$ evolves according the Schrodinger equation
$i(d/dt)|\psi(t)\rangle = H(t) |\psi(t)\rangle$; the Mukunda--Simon phase is defined as \cite{Mukunda}:
$
\Phi ( t) = \arg \langle \psi(0 )| \psi(t )\rangle - \Im\int_{0}^{t}\langle \psi(t^{\prime})|\dot{\psi}(t^{\prime})\rangle dt^{\prime}\,,$
where the dot denotes the derivative with respect to $t^{\prime}$.
The generalization of the above phase to the case of a system with a diagonalizable non-Hermitian Hamiltonian $\mathcal{H}_{NH}(t)$  is  presented in Ref.\cite{Mukunda}.
Its extension to the biorthonormal basis formalism is the following. Denoting with $|\psi_{NH}(t)\rangle$ and $|\widetilde{\psi}_{NH}(t)\rangle $
the solution to the Schrodinger equation $i(d/dt)|\psi_{NH}(t)\rangle = \mathcal{H}_{NH}(t) |\psi_{NH}(t)\rangle$ and to its adjoint equation   $i(d/dt)|\widetilde{\psi}_{NH}(t)\rangle = \mathcal{H}_{NH}^{\dag}(t) |\widetilde{\psi}_{NH}(t)\rangle$,  respectively, the Mukunda--Simon phase is given by:
$
\Phi _{NH}(t) =  \arg \langle \widetilde{\psi}_{NH}(0 )| \psi_{NH}(t )\rangle\ -  \Im \int_{0}^{t} \langle\widetilde{\psi}_{NH}(t^{\prime})| \dot{\psi}_{NH}(t^{\prime}) \rangle dt^{\prime}\,.
$
Such a quantity is reparametrization invariant and it is invariant under  the complex gauge transformations $| \psi_{NH}(t )\rangle \rightarrow  | \psi^{\prime}_{NH}(t )\rangle=  S^{-1}(t) | \psi_{NH}(t )\rangle$ and $ |\widetilde{\psi}_{NH}(t) \rangle  \rightarrow  |\widetilde{\psi}^{\prime}_{NH}(t) \rangle = W^{-1}(t) |\widetilde{\psi}_{NH}(t) \rangle $, where $ W (t) = [S^{-1}(t)]^{\dag}$.   Therefore, $\Phi _{NH}(t)$     is a geometric phase associated with the evolution of a quantum non-Hermitian system.

In the particular case  of  mixed meson systems $M^{0}-\bar{M}^{0}$ one has the  following phases:
\bea\label{FMM}
\Phi_{ M^{0} M^{0}}(  t) & = & \arg \langle \widetilde{M^{0}}(0 )| M^{0}(t )\rangle\, - \Im \int_{0}^{t}  \langle \widetilde{M^{0}}(t^{\prime})| \dot{M}^{0}(t^{\prime})\rangle  dt^{\prime}\,,
\\\label{FMbMb}
\Phi _{\bar{M}^{0}\bar{M}^{0}}(  t) & = & \arg \langle \widetilde{\bar{M}^{0}}(0 )| \bar{M}^{0}(t )\rangle\, - \Im \int_{0}^{t}  \langle \widetilde{\bar{M}^{0}}(t^{\prime})| \dot{\bar{M}}^{0}(t^{\prime})\rangle  dt^{\prime}\,,
\eea
  \bea\label{FM-Mb}
\Phi_{ M^{0} \bar{M}^{0}}(  t) & = & \arg \langle \widetilde{M^{0}}(0 )| \bar{M}^{0}(t )\rangle\, - \Im \int_{0}^{t}  \langle\widetilde{M^{0}}(t^{\prime})| \dot{\bar{M}}^{0}(t^{\prime})\rangle  dt^{\prime}\,,
\\\label{FMb-M}
\Phi _{\bar{M}^{0}M^{0}}(  t) & = & \arg \langle \widetilde{\bar{M}^{0}}(0 )| M^{0}(t )\rangle\, - \Im \int_{0}^{t}  \langle\widetilde{\bar{M}^{0}}(t^{\prime})| \dot{M}^{0}(t^{\prime})\rangle  dt^{\prime}\,.
\eea
$\Phi_{ M^{0} M^{0}}(  t)$ and $\Phi_{ \bar{M}^{0} \bar{M}^{0}}(  t)$ are the phases of the particle $M^{0}$ and of the antiparticle $\bar{M}^{0}$, respectively, and they are  connected to $ CPT$ violation parameter; $\Phi_{ M^{0} \bar{M}^{0}}(  t) $ and $\Phi _{\bar{M}^{0}M^{0}}(  t) $ are the phases due to particle-antiparticle oscillations, and they are linked to $CP$ violation (see below).

Let us analyze  in more detail such phases by starting with $\Phi_{ M^{0} M^{0}}(  t)$ and $\Phi_{ \bar{M}^{0} \bar{M}^{0}}(t)$. By using Eqs.(\ref{B0states1})-(\ref{B0states4}), their explicit form is given by
\bea\label{mucund1B}
\non\Phi _{M^{0} M^{0}}(  t) &=&
 \arg \Big[
e^{ \frac{\Delta \Gamma t}{4}}\lf[(1-\Re z)\cos (m_{L}t)
- \Im z \sin (m_{L}t) \ri]+
e^{- \frac{\Delta \Gamma t}{4}}\lf[(1+\Re z)\cos (m_{H}t) +
\Im z \sin (m_{H}t) \ri]
\\\non
&-& i \lf[e^{ \frac{\Delta \Gamma t}{4}}\lf[(1-\Re z)\sin(m_{L}t) \ri]
+ \Im z \cos (m_{L}t) \ri]+
e^{ -\frac{\Delta \Gamma t}{4}}\lf[(1+\Re z)\sin (m_{H}t) -
 \Im z \cos (m_{H}t)  \ri] \Big]
\\  & + &
\frac{t }{2} \lf(m\,+\,\Delta m\,\Re z\,+\,\frac{  \Delta \Gamma}{2}\,\Im z \ri)\,,
\eea
and
\bea\label{mucundBbar}\non
\Phi _{\bar{M}^{0} \bar{M}^{0}}(  t) &=&
 \arg \Big[
e^{ \frac{\Delta \Gamma t}{4}}\lf[(1+\Re z)\cos (m_{L}t) + \Im z \sin (m_{L}t) \ri]+
e^{- \frac{\Delta \Gamma t}{4}}\lf[(1-\Re z)\cos (m_{H}t) - \Im z \sin (m_{H}t) \ri]
\\\non
&-& i \lf[e^{ \frac{\Delta \Gamma t}{4}}\lf[(1+\Re z)\sin (m_{L}t) - \Im z \cos (m_{L}t) \ri]+
e^{ -\frac{\Delta \Gamma t}{4}}\lf[(1-\Re z)\sin (m_{H}t) + \Im z \cos (m_{H}t) \ri] \ri] \Big]
\\  & + &
\frac{t }{2} \lf(m\,-\,\Delta m\,\Re z\,-\,\frac{ \Delta \Gamma}{2}\,\Im z \ri)\,,
\eea
respectively, where $m = m_{L} + m_{H}$, $\Delta m = m_{H}-m_{L}$ and $\Delta \Gamma = \Gamma_{H}- \Gamma_{L}$.\footnote{The sign of $\Delta \Gamma$ is not yet established for $B$ and $B_s$ mesons, while $\Delta \Gamma < 0$ for $K$ mesons and $\Delta \Gamma > 0$ for $D$ mesons.}
The symbol $\Gamma$  denotes $\Gamma = \Gamma_{L}+\Gamma_{H}$ (Appendix B).
Assuming $\frac{\Delta \Gamma t}{2} $ small,
 which is valid in the range $|\Delta t|< 15 ps$ used in the experimental analysis on $B^{0}-\bar{B}^{0}$ system \cite{Aubert:2004xga}, \cite{Aubert:2007bp},
    Eqs.(\ref{mucund1B}) and (\ref{mucundBbar}) become
\bea\label{mucundB2}
\Phi _{M^{0} M^{0}}(  t) \, \simeq \,
\arg \lf[\cos\lf(\frac{\Delta m t}{2}\ri) + (\Im z - i \Re z )\sin \lf(\frac{\Delta m t}{2}\ri) \ri]
\, + \,
\frac{t }{2} \lf(\Delta m\,\Re z\,+\,\frac{ \Delta \Gamma}{2}\,\Im z \ri)\,,
\eea
and
\bea\label{mucundBbar2}
\Phi_{\bar{M}^{0} \bar{M}^{0}}(  t) \, \simeq \,
\arg  \lf[\cos\lf(\frac{\Delta m t}{2}\ri) - (\Im z - i \Re z )\sin \lf(\frac{\Delta m t}{2}\ri) \ri]
\,- \,
\frac{t }{2} \lf( \Delta m\,\Re z\,+\,\frac{ \Delta \Gamma}{2}\,\Im z \ri)\,,
\eea
respectively.
 These equations show the dependence of the phases on the real and imaginary part of the $z$ parameter defined in Eq.(\ref{zeta}). In particular, the difference between $\Phi _{M^{0} M^{0}}(  t)$ and $\Phi_{\bar{M}^{0} \bar{M}^{0}}(  t) $: $\Delta \Phi (t) = \Phi _{M^{0} M^{0}}(  t) -\Phi_{\bar{M}^{0} \bar{M}^{0}}(  t) $ is due to terms related to $z$ and it is non--zero only in the presence of $CPT$ violation.
Indeed, in the case of $CPT$ invariance, $z=0$,  one has
$ \label{equality}
\Phi^{CPT} _{M^{0} M^{0}}(  t) \, = \,\Phi^{CPT}_{\bar{M}^{0} \bar{M}^{0}}(  t)\,= \,\arg \lf[ \cos\lf(\frac{\Delta m t}{2}\ri)\ri]\,,
$
which is trivially equal to $0$ or $\pi$ and $\Delta \Phi(t) = 0$.

Coming back now to  the phases $\Phi_{M^{0} \bar{M}^{0}}(t)$ and $\Phi _{ \bar{M}^{0}  M^{0}}(  t)$, their explicit expressions are
\bea\label{mucundBBbar}
\Phi_{M^{0} \bar{M}^{0}}(t)\,=\,\frac{\pi}{2}\,-\,\frac{m}{2}t\,+\,\arg \lf[\frac{p}{q}\sqrt{1-z^{2}} \sin \lf[\lf( \Delta m - \frac{i \Delta \Gamma}{2} \ri)\frac{t}{2}\ri]\ri]\,+\,\Im \lf[i\, \frac{p}{q}\,\sqrt{1-z^{2}}\, \lf(  \Delta m - \frac{i \Delta \Gamma}{2} \ri) \frac{t}{2}\ri]\,,
\eea
and
\bea\label{mucundBbarB}
\Phi_{\bar{M}^{0} M^{0} }(t)\,=\,\frac{\pi}{2}\,-\,\frac{m}{2}t\,+\,\arg \lf[\frac{q}{p}\sqrt{1-z^{2}} \sin \lf[\lf( \Delta m - \frac{i \Delta \Gamma}{2} \ri)\frac{t}{2}\ri]\ri]\,+\,\Im \lf[i\, \frac{q}{p}\,\sqrt{1-z^{2}}\, \lf(  \Delta m - \frac{i \Delta \Gamma}{2} \ri) \frac{t}{2}\ri]\,.
\eea
For $\frac{\Delta \Gamma t}{2} \ll 1$  and omitting second order terms in $z$,   one obtains
\bea\label{mucundBBbar2}
\Phi_{M^{0} \bar{M}^{0}}(t)\,=\,\frac{\pi}{2}\,-\,\frac{m t}{2}\,+\,\arg \lf[\frac{p}{q}  \sin \lf(    \frac{\Delta m t}{2}\ri)\ri]\,-\,\frac{\Delta m t}{2}\,  \Re \lf(\frac{p}{q} \ri)\,-\,\frac{\Delta \Gamma t}{2} \, \Im \lf(\frac{p}{q} \ri) \,,
\eea
and
\bea\label{mucundBbarB2}
\Phi_{\bar{M}^{0} M^{0} }(t)\,=\,\frac{\pi}{2}\,-\,\frac{m t}{2}\,+\,\arg \lf[\frac{q}{p}  \sin \lf(    \frac{\Delta m t}{2}\ri)\ri]\,-\,\frac{\Delta m t}{2} \, \Re \lf(\frac{q}{p} \ri)\,-\,\frac{\Delta \Gamma t}{2} \, \Im \lf(\frac{q}{p} \ri)\,,
\eea
and the phase difference is $\Phi_{M^{0} \bar{M}^{0}}(t) - \Phi_{\bar{M}^{0} M^{0} }(t) \neq 0$.
On the contrary, in  the case of $CP$ conservation one has
\bea\label{mucundCP}
\Phi^{CP}_{M^{0} \bar{M}^{0}}(t)\,=\,\Phi^{CP}_{\bar{M}^{0} M^{0} }(t)\,=\,\frac{\pi}{2}\,-\,(m  + \Delta m)\frac{ t}{2}\,+\,\arg \lf[\sin \lf(\frac{\Delta m t}{2}\ri)\ri]\,,
\eea
and there is no phase difference.

\emph{Numerical analysis:}
The features of the phases related to the parameter $z$, $\Phi_{M^{0} M^{0}}$, $\Phi_{\bar{M}^{0} \bar{M}^{0}}$ and $\Delta \Phi$ are analyzed in detail for the $B_{s}$   system.
Such a system is particulary appropriate for the study of non-cyclic phases since many particle oscillations occur within its  lifetime. Another useful system for such analysis is the neutral kaon one \cite{Di Domenico}.

For the $B_{s}$ mesons, one takes $m_{s} = 1.63007 \times 10^{13} ps^{-1}$,
$\Delta m_{s} = 17.77 ps^{-1}$,
$ \Gamma_{s} = 0.678 ps^{-1}$,
$\Delta \Gamma_{s}  = -0.062 ps^{-1}$.
Moreover,  one considers values of $\Re z$ and $\Im z$ in the intervals: $-0.1\leq \Re z \leq 0.1$ and $-0.1\leq \Im z \leq 0.1 $ which are consistent with the experimental data \cite{Aubert:2007bp}. Notice that, in such intervals for  $\Re z$ and $\Im z$,
the phases $\Phi_{ B_{s}^{0} B_{s}^{0}} $, $\Phi_{ \bar{B}_{s}^{0} \bar{B}_{s}^{0}} $ and $\Delta \Phi$ are weakly depending  on the value of $\Im z $. Indeed, for example, in the time interval of the $B^{0}_{s}$ life time, the shape variation of  $\Phi_{ B_{s}^{0} B_{s}^{0}}$ and
$\Phi_{ \bar{B}_{s}^{0} \bar{B}_{s}^{0}} $ with $\Im z $ is at most of the order of $0.2\%$, so that one can fix an arbitrary value of $\Im z $ in the values interval  $[-0.1, 0.1]$ and study the non-cyclic phases as functions of  time for different values  of $\Re z   $.
 In Figs. $(1)$, $(2)$ and $(3)$ the phases are drawn   for $\Im z = 0 $. In order to  better show the behavior of the phases, the figures contain two plots A) and B)  of the same phase for sample values of $\Re z$ belonging to the intervals  $ [-0.1, 0]$ and $ [0, 0.1]$, respectively.

\begin{figure}
\begin{picture}(300,220)(0,0)
\put(-90,20){\resizebox{9 cm}{!}{\includegraphics{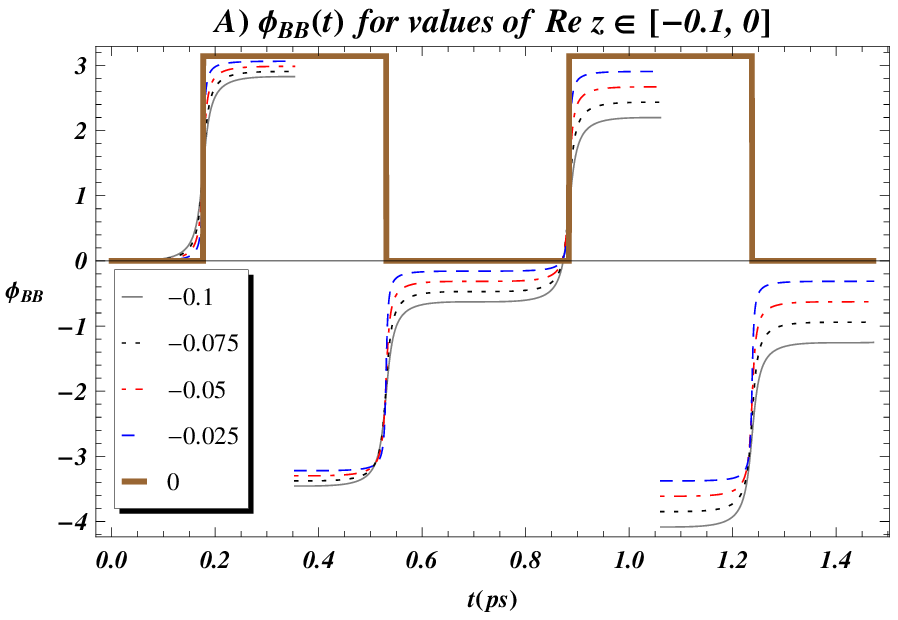}}}
\put(160,20){\resizebox{9 cm}{!}{\includegraphics{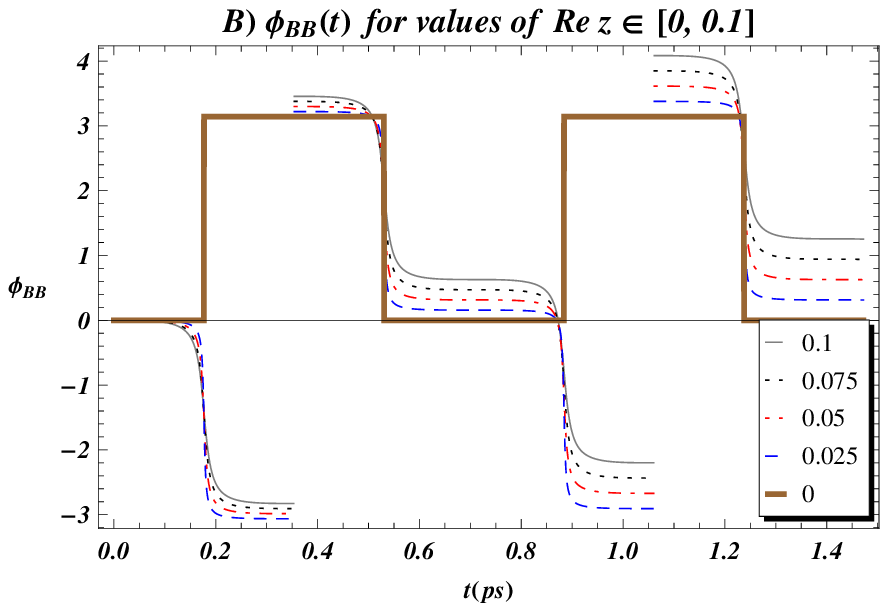}}}
\end{picture}
\caption{\em Plots of $\Phi_{ B_{s}^{0} B_{s}^{0}} $ as a function of time $t$ for   $\Im z = 0$ and different values of $\Re z$. In  picture A) $\Phi_{ B_{s}^{0} B_{s}^{0}}(t) $ is reported for sample values of $Re z \in [-0.1, 0]$ as indicated in the inset. In picture B) $\Phi_{ B_{s}^{0} B_{s}^{0}}(t) $ is plotted for sample values of $Re z \in [0, 0.1]$ as shown in the inset.}
\label{pdf}
\end{figure}

\begin{figure}
\begin{picture}(300,220)(0,0)
\put(-90,20){\resizebox{9 cm}{!}{\includegraphics{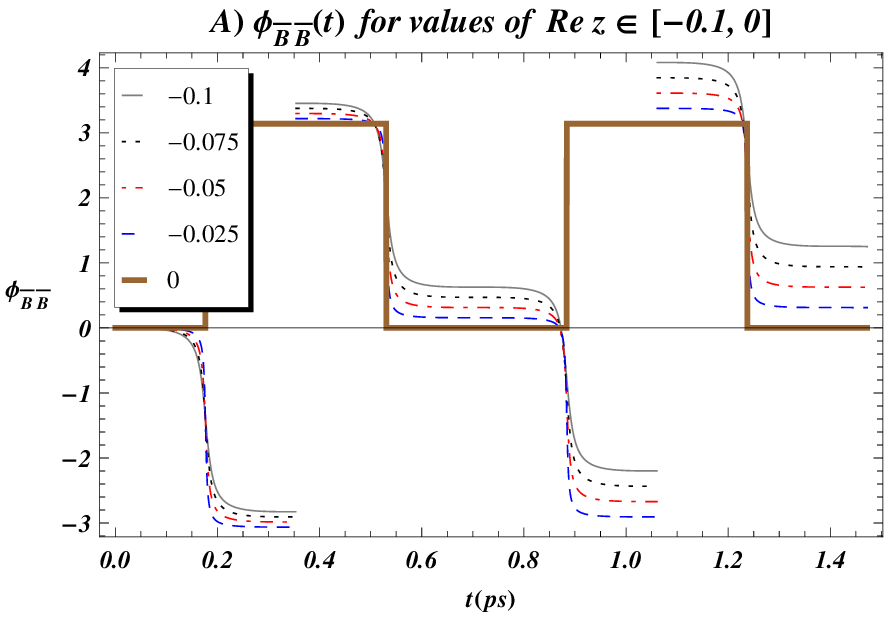}}}
\put(160,20){\resizebox{9 cm}{!}{\includegraphics{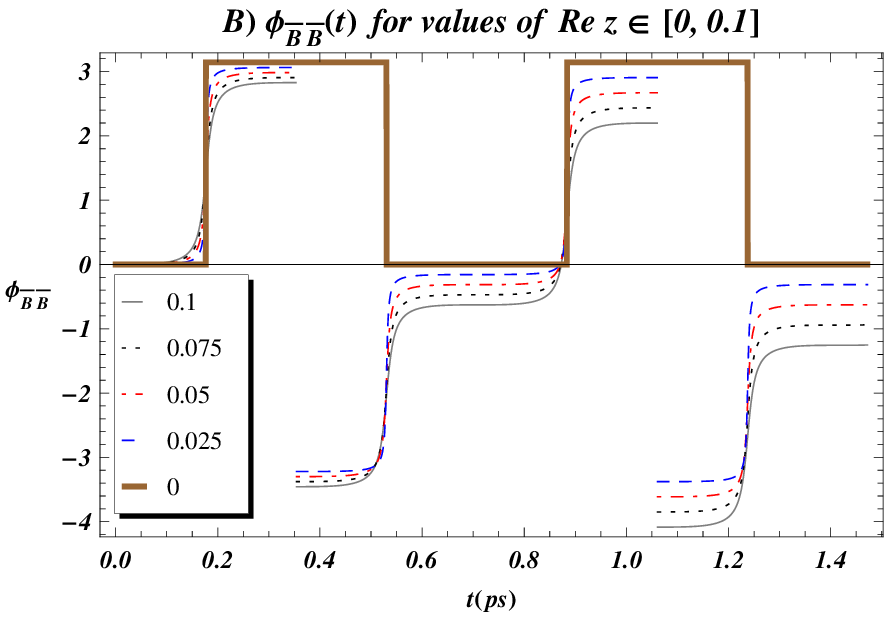}}}
\end{picture}
\caption{\em Plots of $\Phi_{ \bar{B}_{s}^{0} \bar{B}_{s}^{0}} $ as a function of time for  $\Im z = 0$ and different sample values of $\Re z$ as in Figure 1.  }
\label{pdf}
\end{figure}

\begin{figure}
\begin{picture}(300,220)(0,0)
\put(-90,20){\resizebox{9 cm}{!}{\includegraphics{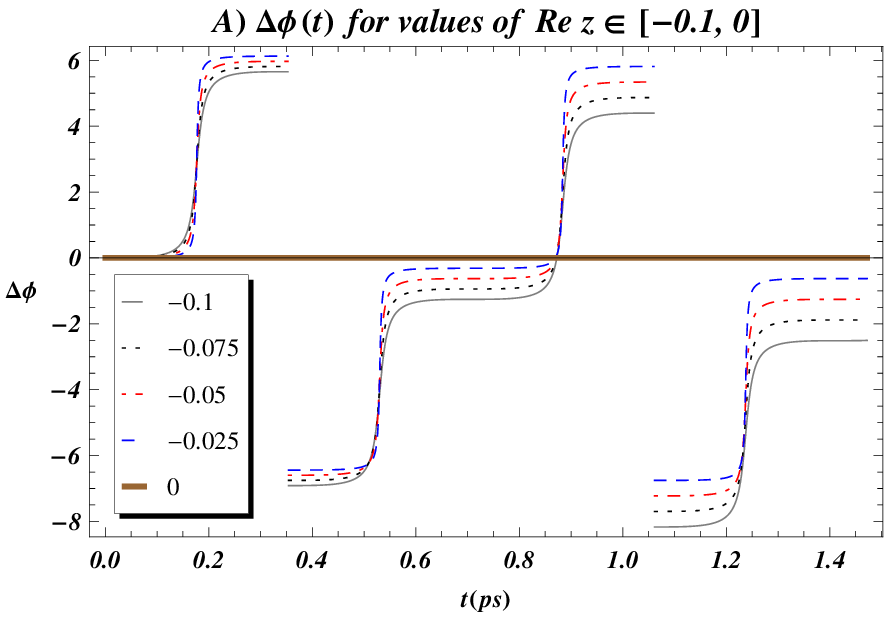}}}
\put(160,20){\resizebox{9 cm}{!}{\includegraphics{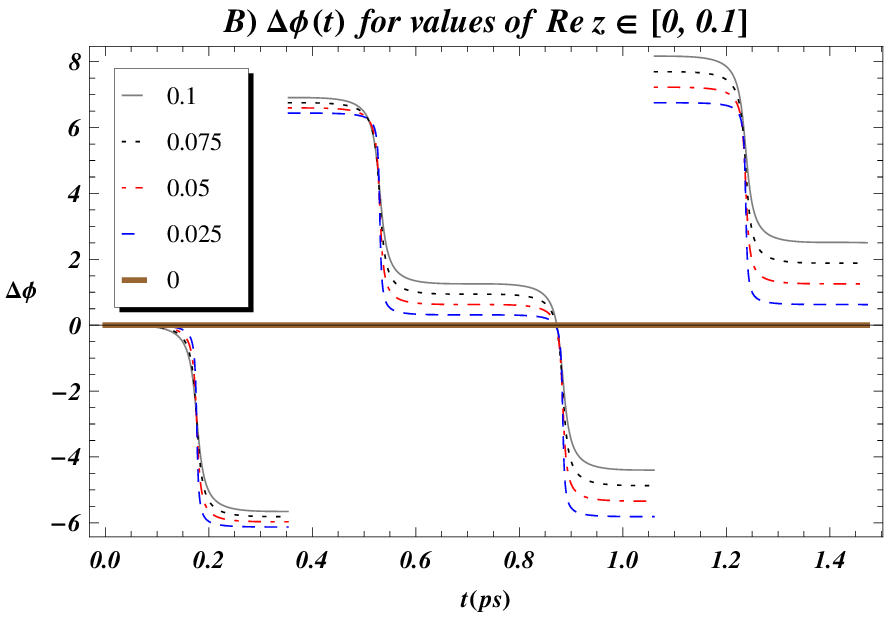}}}
\end{picture}
\caption{\em Plots of $\Delta \Phi $ as a function of time for $\Im z = 0 $ and different sample values of $\Re z$ as in Figures 1 and 2. }
\label{pdf}
\end{figure}

The plots show that in the case of $CPT$ violation, the phases $\Phi_{ B_{s}^{0} B_{s}^{0}} $ and $\Phi_{ \bar{B}_{s}^{0} \bar{B}_{s}^{0}} $ are clearly non-trivial, in fact they can assume values different from $0$ and $\pi$ and, in particular, the phase difference $\Delta \Phi$ is non--zero.

\section{Conclusions }

In the presence of $T$ violation the effective Hamiltonian of mixed  meson systems is non-Hermitian and non-normal. The left and right eigenstates of  $\mathcal{H}$ are independent sets of vectors
that are not connected by complex conjugation. Then $\mathcal{H}$ cannot be diagonalized by a single unitary transformation but by a complete biorthonormal set of vectors.

The correct flavor states are then derived by using the biorthonormal basis formalism. They are used to compute the non-cyclic phases for oscillating  mesons and  the asymmetries describing the $CP$ and $CPT$ violations (see Appendix B). The obtained asymmetries are equivalent to the ones achieved by the usual formalism.

The main outcome of the present work is the study of the  Mukunda Simon phases for mixed mesons and the discovery of the fact that the geometric phases appearing in the evolution of the meson  $\Phi_{M^{0}  {M}^{0}}(t)$ and of its antiparticle $\Phi_{\bar{M}^{0} \bar{M}^{0} }(t)$ depend on the $CPT$ violating parameter  $z$.
 In particular, only in the case of $CPT$ symmetry breaking, such phases are non trivial and the phase difference $\Delta \Phi$ between particle and antiparticle is non-zero.

The possibility  of $CPT$  violation has been investigated in detail by analyzing the  Mukunda Simon phases for the neutral $B_{s}$ system. Such a system, together with the kaon, is especially suitable for the study of geometric phases.
The high precision of the upcoming experiments on the  $B_{s}^{0}$ mesons will allow us, in the next future, to completely determine  the dynamics of such particles, thus such experiments and the ones analyzing kaons dynamics might allow an accurate analysis of the geometric phases and in particular a measurement of the phase difference  $\Delta \Phi$ generated in the time evolution of the particle and the antiparticle.
  Such measurements might represent a completely alternative method to test one of the most important symmetries of the nature.

Finally, it has been also shown that the Mukunda-Simon phases $\Phi_{M^{0} \bar{M}^{0}}(t)$, $\Phi_{\bar{M}^{0} M^{0} }(t)$ and the Aharonov Anandan invariants $s_{ M^{0}\bar{M}^{0}}( t)$, $s_{\bar{M}^{0} M^{0}}( t)$
(see Appendix C) due to meson oscillations are related to the  $CP$ violation parameters.
 $ CPT$ violation should not affect these phases, as the corrections
are quadratic and expected to be negligible for small $z$. Thus, a  study
of the non-cyclic phases might be useful
also for the analysis of the $ CP$ symmetry breaking.

\section*{Acknowledgements}

Partial financial support from Miur is acknowledged.

\appendix

\section{Other expressions of  the states $|M^{0}(t)\rangle$ and $|\bar{M}^{0}(t)\rangle$ }

The mass eigenstates $|M_{L}\rangle$  and $| M_{H}\rangle$  are written in terms of $|M^{0}\rangle$,
$|\bar{M}^{0}\rangle$ as \footnote{ note that  $\langle M_{j} |M_{i \neq j}\rangle \neq 0 \,;$ for example, in the kaon case, one has $\langle K_{S} |K_{L}\rangle \neq 0$.}
\bea\label{Pa1}
|M_{L}\rangle &=& p_{L} \,|M^{0} \rangle\,+\, q_{L}\, |\bar{M}^{0} \rangle \,,
\\\label{Pa2}
| M_{H}\rangle &=& p_{H} \,|M^{0} \rangle\,-\, q_{H}\, |\bar{M}^{0} \rangle \,,
\eea
and, in a similar way, $\langle \widetilde{M}_{L}|$  and $\langle \widetilde{M}_{H}|$  are expressed as
\bea
\label{Pa3}
\langle\widetilde{M}_{L} |  &=& \frac{1}{q_{L} p_{H}+q_{H} p_{L}} \lf[q_{H}\,\langle \widetilde{M^{0}}|  \,+\,
p_{H}\,\langle \widetilde{ \bar{M}^{0}}| \ri]\,,
\\\label{Pa4}
\langle\widetilde{M}_{H}| &=& \frac{1}{q_{L} p_{H}+q_{H} p_{L}} \lf[q_{L}\,\langle \widetilde{M^{0}}|  \,-\,
p_{L}\,\langle \widetilde{ \bar{M}^{0}}| \ri]\,.
\eea
Then at time $t$, the states $|M^{0}(t)\rangle$ and $|\bar{M}^{0}(t)\rangle$ in terms of $|M_{L}\rangle$ and
$|M_{H}\rangle$ are:
\bea\label{P0(t)}
|M^{0}(t)\rangle &=& \frac{1}{q_{L} p_{H}+q_{H} p_{L}} \lf[q_{H}\,|M_{L} \rangle\, e^{-i \lambda_{L} t}\,+\,
q_{L}\,|M_{H} \rangle\, e^{-i \lambda_{H} t}  \ri]\,,
\\
|\bar{M}^{0}(t)\rangle &=& \frac{1}{q_{L} p_{H}+q_{H} p_{L}} \lf[p_{H}\,|M_{L} \rangle\, e^{-i \lambda_{L} t}\,-\,
p_{L}\,|M_{H} \rangle\, e^{-i \lambda_{H} t}  \ri]\,,
\\
\langle\widetilde{M^{0}}(t)|  &=&  p_{L}\, \langle \widetilde{M}_{L}| \, e^{ i \lambda_{L} t}\,+\,
p_{H}\,\langle \widetilde{M}_{H}| \, e^{ i \lambda_{H} t} \,,
\\\label{BP0(t)}
\langle\widetilde{ \bar{M}^{0}}(t)| &=& q_{L}\, \langle \widetilde{M}_{L}| \, e^{ i \lambda_{L} t}\,-\,
q_{H}\,\langle \widetilde{M}_{H}| \, e^{ i \lambda_{H} t} \,.
\eea
These states can be expressed also in the bases $\{| M^{0} \rangle\,, |\bar{M}^{0} \rangle \,,  \langle \widetilde{M^{0}}| \,,  \langle \widetilde{\bar{M}^{0}}|\}$ as
\bea\label{B0time1}
|M^{0}(t)\rangle &=& [g_{+}(t)\,+\,z\,g_{-}(t)]\,|M^{0} \rangle \,-\, \sqrt{1-z^{2} }\,\frac{q}{p}\,g_{-}(t)\,|\bar{M}^{0} \rangle\,,
\\
\label{B0time2}
|\bar{M}^{0}(t)\rangle &=& -\, \sqrt{1-z^{2} }\,\frac{p}{q}\,g_{-}(t)\,|M^{0} \rangle\,+\,[g_{+}(t)\,-\,z\,g_{-}(t)]\,|\bar{M}^{0} \rangle \,,
\\
\label{B0time3}
\langle \widetilde{M^{0}}(t)| &=& [\widetilde{g}_{+}(t)\,+\,z\,\widetilde{g}_{-}(t)]\,\langle \widetilde{M^{0}}|\,-
\,\sqrt{1-z^{2} }\,\frac{p}{q}\,\widetilde{g}_{-}(t)\,\langle \widetilde{\bar{M}^{0}}| \,,
\\
\label{B0time4}
\langle \widetilde{\bar{M}^{0}}(t)| &=& -
\,\sqrt{1-z^{2} }\,\frac{q}{p}\,\widetilde{g}_{-}(t)\,\langle \widetilde{M^{0}}|\,+\,[\widetilde{g}_{+}(t)\,-\,z\,\widetilde{g}_{-}(t)]\,\langle \widetilde{\bar{M}^{0}}| \,,
\eea
with $g_{\mp}(t) = \frac{1}{2}(e^{-i\lambda_{H}t}\mp e^{-i\lambda_{L}t})$
and $\widetilde{g}_{\mp}(t) = \frac{1}{2}(e^{ i\lambda_{H}t}\mp e^{ i\lambda_{L}t})$.

\section{CP and CPT asymmetries}

The expressions of the asymmetries $A_{T}$ and $A_{CPT}$  describing a departure from time reversal
and  $CPT$ invariances, respectively, are calculated by using the states derived in the biorthonormal formalism, Eqs.(\ref{B0states1})-(\ref{B0states4}).
The obtained  results are equivalent to the asymmetries computed by using the usual formalism  \cite{Aubert:2007bp}, \cite{Lenz:2010gu}.

Let us begin by computing the $T$ asymmetry.
The violation of time reversal invariance can be revealed by the comparison between the probability
of transition from $\bar{M}^{0}$ to $M^{0}$, $P_{\bar{M}^{0}\rightarrow M^{0}}$, and the probability
of transition
from $M^{0}$ to $\bar{M}^{0}$, $P_{M^{0}\rightarrow \bar{M}^{0}}$, in the asymmetry:
\bea
A_{T}(\Delta t)\,=\,\frac{P_{\bar{M}^{0}\rightarrow M^{0}}(\Delta t)-P_{M^{0}\rightarrow \bar{M}^{0}}(\Delta t)}
{P_{\bar{M}^{0}\rightarrow M^{0}}(\Delta t)+P_{M^{0}\rightarrow \bar{M}^{0}}(\Delta t)}\,
\eea
with $\Delta t = t_{f}- t_{i}$ denoting the time interval between the initial time $t_{i}$ and the final time $t_{f}$. The transition amplitudes
$A_{\bar{M}^{0}\rightarrow M^{0}}(\Delta t)$ and
$A_{M^{0}\rightarrow \bar{M}^{0}}(\Delta t)$ are given respectively by
\bea\label{Amb-m}
A_{\bar{M}^{0}\rightarrow M^{0}}(\Delta t)&=&  \langle\widetilde{M^{0}}(t_{f})|\bar{M}^{0}(t_{i})\rangle\,=\,\langle\widetilde{M^{0}}|e^{-i H \Delta t}|\bar{M}^{0}\rangle
\,=\,\frac{1}{2} \frac{q}{p}\sqrt{1-z^{2} } \lf
(e^{- i \lambda_{L} \Delta t} - e^{- i \lambda_{H} \Delta t} \ri)\,,
\\\label{Am-mb}
A_{M^{0}\rightarrow \bar{M}^{0}}(\Delta t)&=& \langle\widetilde{ \bar{M}^{0}}(t_{f})|M^{0}(t_{i})\rangle
\,=\, \langle\widetilde{ \bar{M}^{0}}|e^{-i H \Delta t}|M^{0} \rangle
\,=\,\frac{1}{2} \frac{p}{q}\sqrt{1-z^{2} } \lf
(e^{- i \lambda_{L} \Delta t} - e^{- i \lambda_{H} \Delta t} \ri)\,.
\eea
The results in Eqs.({\ref{Amb-m})-(\ref{Am-mb}) are obtained by introducing the identity operator $|M_{L}\rangle \langle \widetilde{M_{L}}|
+ |M_{H}\rangle \langle \widetilde{M_{H}}| = 1$ on the right side of the operator $e^{-i H \Delta t}$. The corresponding transition probabilities are then
\bea
P_{\bar{M}^{0}\rightarrow M^{0}}(\Delta t)&=& \lf| \langle\widetilde{M^{0}}(t_{f})|\bar{M}^{0}(t_{i})\rangle\ri|^{2}\,=
 \frac{1}{2} \lf| \frac{q}{p}\ri|^{2} \lf| \sqrt{1-z^{2} } \ri|^{2} e^{- \frac{\Gamma}{2} \Delta t} \lf[ \cosh \lf( \frac{\Delta\Gamma  \Delta t}{2}\ri) -  \cos (\Delta m  \Delta t)
 \ri]\,,
\\
P_{M^{0}\rightarrow \bar{M}^{0}}(\Delta t)&=& \lf|\langle\widetilde{ \bar{M}^{0}}(t_{f})|M^{0}(t_{i})\rangle\ri|^{2}\,=
\frac{1}{2} \lf| \frac{p}{q}\ri|^{2} \lf| \sqrt{1-z^{2} } \ri|^{2} e^{- \frac{\Gamma}{2} \Delta t} \lf[ \cosh \lf( \frac{\Delta\Gamma  \Delta t}{2}\ri) -  \cos (\Delta m  \Delta t)
 \ri]\,.
\eea
  The asymmetry $A_{T}$ is   time independent and it is given by
\bea\label{A(T)}
A_{T}\,=\,\frac{1- \lf|\frac{q}{p}\ri|^{4}}{1+ \lf|\frac{q}{p}\ri|^{4}}\,.
\eea
 A value different from zero of the quantity in Eq(\ref{A(T)})
indicates a direct $T$ violation independent from $CPT$ violation. The result (\ref{A(T)}) coincides with the Eq.(54) in Ref.\cite{Lenz:2010gu}.

In a similar way, the violation of $CPT$ invariance can be revealed by the comparison between the probability
of transition from $M^{0}$ to $ M^{0}$, $P_{M^{0}\rightarrow M^{0}}$,  and the probability
of transition
from  $\bar{M}^{0}$ to $\bar{M}^{0}$, $P_{\bar{M}^{0}\rightarrow \bar{M}^{0}}$, in the asymmetry
\bea\label{A(CPT)}
A_{CPT}(\Delta t)\,=\,\frac{P_{M^{0}\rightarrow M^{0}}(\Delta t) - P_{\bar{M}^{0}\rightarrow \bar{M}^{0}}(\Delta t)}
{P_{M^{0}\rightarrow M^{0}}(\Delta t) + P_{\bar{M}^{0}\rightarrow \bar{M}^{0}}(\Delta t)}\,.
\eea
The transition amplitudes $A_{M^{0}\rightarrow M^{0}}(\Delta t)$
and
$A_{\bar{M}^{0}\rightarrow \bar{M}^{0}}(\Delta t)$ are given respectively by
\bea
\label{AM0}
A_{M^{0}\rightarrow M^{0}}(\Delta t)&=& \langle\widetilde{ M^{0}}(t_{f})|M^{0}(t_{i})\rangle
\,=\, \langle\widetilde{ M^{0}}|e^{-i H \Delta t}|M^{0} \rangle
\,=\,\lf(\frac{1+z}{2}\ri)\,
e^{- i \lambda_{H} \Delta t} + \lf(\frac{1-z}{2}\ri)\, e^{- i \lambda_{L} \Delta t} \,,
\\\label{AM0-bar}
A_{\bar{M}^{0}\rightarrow \bar{M}^{0}}(\Delta t)&=&  \langle\widetilde{\bar{M}^{0}}(t_{f})|\bar{M}^{0}(t_{i})\rangle
\,=\, \langle\widetilde{\bar{M}^{0}}|e^{-i H \Delta t}|\bar{M}^{0} \rangle
\,=\,\lf(\frac{1-z}{2}\ri)\,
e^{- i \lambda_{H} \Delta t} + \lf(\frac{1+z}{2}\ri)\, e^{- i \lambda_{L} \Delta t} \,,
\eea
where again the relation $|M_{L}\rangle \langle \widetilde{M_{L}}|
+ |M_{H}\rangle \langle \widetilde{M_{H}}| = 1$ has been introduced on the right side of $e^{-i H \Delta t}$.
The corresponding transition probabilities are then
\bea \non
 && P_{M^{0}\rightarrow M^{0}}(\Delta t) = \lf|\langle\widetilde{ M^{0}}(t_{f})|M^{0}(t_{i})\rangle \ri|^{2}
\\
 && =
 e^{- \frac{\Gamma}{2}  \Delta t} \lf[\lf(\frac{1+|z|^{2}}{2}\ri) \cosh \lf(\frac{\Delta \Gamma \Delta t}{2} \ri) - \Re z \sinh \lf(\frac{\Delta \Gamma \Delta t}{2} \ri) + \lf(\frac{1-|z|^{2}}{2}\ri)
\cos (\Delta m \Delta t) + \Im z  \sin (\Delta m \Delta t) \ri] \,,
\\ \non
&&P_{\bar{M}^{0}\rightarrow \bar{M}^{0}}(\Delta t)= \lf|\langle\widetilde{\bar{M}^{0}}(t_{f})|\bar{M}^{0}(t_{i})\rangle\ri|^{2}
\\
&&=
e^{- \frac{\Gamma}{2}  \Delta t} \lf[\lf(\frac{1+|z|^{2}}{2}\ri) \cosh \lf(\frac{\Delta \Gamma \Delta t}{2} \ri) + \Re z \sinh \lf(\frac{\Delta \Gamma \Delta t}{2} \ri) + \lf(\frac{1-|z|^{2}}{2}\ri)
\cos (\Delta m \Delta t) - \Im z  \sin (\Delta m \Delta t) \ri] \,.
\eea
 The asymmetry $A_{CPT}$ is thus given by
\bea\label{A(CPT)B0}
A_{CPT}(\Delta t)\,= \, \frac{-2\,\Re z\,\sinh \lf(\frac{\Delta \Gamma \Delta t}{2} \ri)\, + \,2\,  \Im z\, \sin (\Delta m \Delta t)}
{(1+ |z|^{2})\, \cosh \lf(\frac{\Delta \Gamma \Delta t}{2} \ri)\, +  \,(1- |z|^{2})\,   \cos (\Delta m \Delta t) }\,.
\eea
Omitting second order terms in $z$ and making the approximation
  $\sinh \lf(\frac{\Delta \Gamma \Delta t}{2} \ri)\simeq \frac{\Delta \Gamma \Delta t}{2}$ which is valid in the range $|\Delta t|< 15 ps$ used in the experimental analysis of the $B^{0}-\bar{B}^{0}$ systems \cite{Aubert:2006nf}, \cite{Aubert:2007bp}, one has
\bea\label{A(CPT)B0-1}
A_{CPT}(\Delta t)\,\simeq \, \frac{ -\Re z\,   \Delta \Gamma \Delta t  \, + \,2\,  \Im z\, \sin (\Delta m \Delta t)}
{  \cosh \lf(\frac{\Delta \Gamma \Delta t}{2} \ri)\, +  \,  \cos (\Delta m \Delta t) }\,,
\eea
which coincides with Eq.(6) of Ref.\cite{Aubert:2007bp}.
In the case of $CPT$ invariance, $z =0$ and $A_{CPT}=0$.

\section{ Aharonov--Anandan phase and $CP$ violation}

The Aharonov--Anandan invariant  is defined as  \cite{Aharonov:1987gg}
$
s= 2 \int_{0}^{  t}  \Delta E (t^{\prime}) \, dt^{\prime}\,,
$
where $\Delta E $ is the variance of the energy $E$.
The generalization to systems with a non-Hermitian Hamiltonian is presented in Ref.\cite{Nesterov} where the biorthonormal basis formalism is also used. For a system with a complete biorthonormal basis
$ \{|\psi(t)\rangle ,  \langle\widetilde\psi(t)| \}$, the variance is given by
$
\Delta E_{NH}^{2}(t) = \langle\widetilde\psi(t)|H^{2}|\psi(t)\rangle -  (\langle\widetilde\psi(t)|H|\psi(t)\rangle)^{2},
$
and  the Aharonov--Anandan phase is
$
s_{NH}= 2 \int_{0}^{  t}  |\Delta E_{NH} (t^{\prime})| \, dt^{\prime}\,
$.
In the  particular case of the mixed meson systems, one has the following variances:
\bea\label{e1}
\Delta E_{M^{0} {M}^{0}} ( t ) & =&  \Delta E_{\bar{M}^{0}\bar{M}^{0}} ( t ) \,=\,
\frac{1}{2} \sqrt{(1-z^{2})}\,(\lambda_{H}- \lambda_{L})\, ,
\\\label{e2}
\Delta E_{M^{0}\bar{M}^{0}} ( t ) & =& \langle\widetilde{M^{0}} ( t)|H|\bar{M}^{0}(  t)\rangle\,=\,
-\frac{p}{q}\frac{\sqrt{(1-z^{2})}}{2} (\lambda_{H}- \lambda_{L})\,,
\\\label{e3}
\Delta E_{M^{0}\bar{M}^{0}} ( t ) & =& \langle\widetilde{\bar{M}^{0}} ( t)|H|M^{0}(  t)\rangle\,=\,
-\frac{q}{p}\frac{\sqrt{(1-z^{2})}}{2} (\lambda_{H}- \lambda_{L})\,.
\eea
Such relations show that the variances depend on $z^{2}$. Moreover, since    $\Delta E_{M^{0} {M}^{0}} ( t ) = \Delta E_{\bar{M}^{0}\bar{M}^{0}} ( t ) $, then the corresponding invariants for particle and antiparticle are equal.  These facts mean that  Aharonov--Anandan invariants do not represent a good tool to test $CPT$ invariance. However, such phases could be useful in the study of $CP$ violation. Indeed, by neglecting the second order dependence on the $z$ parameter, one has:
  \bea\label{s1}
s_{M^{0}M^{0}}( t)&=& s_{\bar{M}^{0}\bar{M}^{0}}( t)\,=\, 2 \int_{0}^{ t} |\Delta E_{M^{0}M^{0}} (  t^{\prime} )|\, dt^{\prime}
=  \sqrt{(\Delta m)^{2}+\frac{(\Delta \Gamma)^{2}}{4} }\, t \,,
\\\label{s2}
s_{ M^{0}\bar{M}^{0}}( t) & =& 2 \int_{0}^{ t} |\Delta E_{M^{0}\bar{M}^{0}} ( t^{\prime} )|\, dt^{\prime} = \lf|\frac{p}{q}\ri|\sqrt{(\Delta m)^{2}+\frac{(\Delta \Gamma)^{2}}{4} }\, t \,,
\\\label{s3}
s_{\bar{M}^{0}M^{0}} ( t) &= & 2 \int_{0}^{   t} |\Delta E_{\bar{M}^{0}M^{0}} ( t^{\prime} )|\, dt^{\prime}  = \lf|\frac{q}{p}\ri|\sqrt{(\Delta m)^{2}+\frac{(\Delta \Gamma)^{2}}{4} }\, t \,.
\eea
 The phase  $s_{ M^{0}\bar{M}^{0}}( t) $ is different from $s_{ \bar{M}^{0}M^{0}}( t) $   because of the $CP$ violation $p \neq q$, independently from $CPT$ violation.
  Eqs.(\ref{s2}) and (\ref{s3})   can be then used to compute the following quantity:
\bea
\frac{s_{ M^{0}\bar{M}^{0}} - s_{\bar{M}^{0}M^{0}} }{ s_{ M^{0}\bar{M}^{0}} + s_{\bar{M}^{0}M^{0}}}=
\frac{|p/q|-|q/p|}{|p/q|+|q/p|}=\frac{|H_{12}|-|H_{21}|}{|H_{12}|+|H_{21}|}\,,
\eea
which coincides with the $CP$ and $T$ violating parameter $\varepsilon$ defined in Eq.(\ref{CPpar}).
Thus, the Aharonov--Anandan phases could represent a completely new way to estimate the parameter $\varepsilon$
  in mixed meson systems such as the $K^{0}-\bar{K}^{0}$ one \cite{Di Domenico}.
In the case of $CP$ conservation one should have $s_{M^{0}M^{0}}( t)\,=\, s_{\bar{M}^{0}\bar{M}^{0}}( t)\,=\,s_{ M^{0}\bar{M}^{0}}( t)\,=\,s_{\bar{M}^{0}M^{0}} ( t)$ and $\varepsilon = 0$.


\end{document}